# Alternative Theory of Nucleation in Super-Cooled Liquids

L.-C. Valdès, F. Affouard

Unité Matériaux et Transformations (UMET), UMR CNRS 8207, UFR de Physique, Bât. P5, Université de Lille 1, 59655 Villeneuve d'Ascq, France

corresponding author: laurent-charles.valdes@univ-lille1.fr

**ABSTRACT**

Having discovered a dimension anomaly in two key formulas of the Classical Nucleation Theory (CNT) but wishing to remain in the style of this theory, we propose to approach nucleation on the basis of the Zeldovich unsteady rate formula, with dimensionally correct expressions for the nucleation rate and time constant.

Beforehand, the problematic status - physical size or parameter - of interfacial tension in CNT was audited. The results of numerical simulations on nucleation in various attraction but fixed repulsion conditions of the interaction potential have led us to motivate, then establish, a thermal evolution law for interfacial tension. Taking into account a scale dependance in the vicinity of absolute zero in temperature, this law is of stretched Arrhenius type. It works with the Zeldovich formula, notably for the determination of adjusting parameters.

A remarkably accurate adjusting of this law to the numerical simulations has been obtained and led us to exhibit melting crystal volume as a measure of potential attractivity. It should allow accurate forecasts of instantaneous nucleation rate and average nucleation duration in physical or numerical super-cooled monoatomic liquids.

**I. INTRODUCTION**

Studies to understand and control the crystallization process led to the concepts of nucleation and growth, to be found at the core of the Classical Nucleation Theory (CNT). This flew off in 1926 in the pioneering work of Volmer and Weber (Ref. 1), was given added impetus by Farkas (Ref. 2) and reached its peak with the significant work of Becker and Döring (Ref. 3), Zeldovich (Ref. 4) and Frenkel (Ref. 5), to just mention the most famous. Today, it is a sophisticated theory with much interweaving between models of all natures, from fundamental physics to empirical, which may contribute to confusing processes and hiding certain fundamental requirements of Physics. Is that the reason why small dimension anomalies could incubate till today in two prominent formulas?

All the same, the CNT has the merit of still providing the elements of a solid understanding of modern treatments of crystallization. It particularly highlights tensions acting at interfaces or interfacial tension. On the molecular scale, the direction and intensity of these tensions appear of purely geometrical nature, their direction resulting from asymmetries between spatial structures on either side of the interface and their intensity from the gap between the attracting centres due to temperature. From a theoretical point of view, interfacial tension is evidently considered as a true physical size. However, from a practical point of view, their experimental determination having stood up to all experimental attempts due to the metastability of super-cooled liquids, it is essentially used to adjust formalism to experimental results. Due to this ambiguous status, at the same time both variable and parameter (Ref. 6), interfacial tension is like a doubtful pivot on which all the theory sits.

Starting from this disturbing point, the initial goal for this work was to address the question of the exact status of interfacial tension in CNT. The first thing to do was, to judge on the evidence, to determine the interfacial tension in the sense of CNT. Investigating nucleation by means of computer Molecular Dynamics simulations allowed the value of all key-sizes of nucleation following CNT to be determined and the interfacial tension, thus reduced to the only unknown, to be easily solved. But the physical relevance of the obtained results not being correctly estimated from a unique thermal evolution of interfacial tension, several different nucleation conditions had, on principle, to be implemented. The choice was made to characterize these nucleation conditions by an *a priori* influent parameter acting at the fundamental level, namely, following the above geometrical point of view and in coherence with the results presented in Ref. 7, by the potential attraction. This choice led us to implement a family of interaction potentials having a fixed repulsive part and a parameterizable attractive part.

By settling the question of the status of interfacial tension in CNT, the physical limits and deficiencies of this theory so appeared that finding a replacement solution to CNT was imperative, as the second goal to this work. Opportunely, combining the solution of the above mentioned dimension anomalies and the Zeldovich formula could constitute the starting point of a new formalism allowing us to keep the characteristics of simplicity and efficiency of CNT.

Part II deals with the initial goal of this study; successively are presented: the interaction potential used to carry out the numerical simulations (Sect. A), the simulation results on which lies all this study (Sect. B), the interfacial tensions following CNT related to different attractions (Sect. C) and comments to be deduced (Sect. D). The last motivated the second goal which occupies part III; a new setting of the problem, in which the dimension anomalies do not appear, is proposed (Sect. E), the general form of the new interfacial tension thermal evolution law is established (Sect. F), its accurate characteristics are determined on the basis of the simulation results (Sect. G) and concluded by some comments (Sect. H). Part IV presents implementations in order to compare forecasts with experiments. Part V gives the conclusions of this study.

## II. ESTIMATION OF TCN BY ITS INTERFACIAL TENSIONS

### A. Interaction potential of the numerical simulations

To allow nucleation conditions to be diversified, the interaction potential implemented in this study was conceived in such a way that the average inter-molecular distance could be imposed, the latter *a priori* having influence on interfacial tension. Then, this potential possesses a fixed and a parameterizable part. Due to the monoatomic nature of simulated particles (Sect. B), it should be as near as possible to the Lennard-Jones potential, known for its excellence in describing the argon fluid behaviour, the mathematical form of which being:

$$E_{LJ}(r) = \varepsilon\left[\left(\frac{r_0}{r}\right)^{12} - 2\left(\frac{r_0}{r}\right)^{6}\right] \quad (1).$$

This simply expresses impenetrability, coherence and fluidity at low density, the fundamental physical characteristics from which all others follow through the complex play of molecular interactions. $r_0$ is the distance at which this potential reaches its minimum and $\varepsilon$ is its minimum value; the relation $r_0 = 2^{1/6}.\sigma$ also exists between $r_0$ and the efficient section $\sigma$ of particles.

The fixed repulsive part of the potential implemented in numerical simulations is that of the Lennard-Jones potential, while the attractive part, which involves the parameterized set of curves described below, is compelled to respect its general shape, the goal being to achieve a correct description of fundamental characteristics of matter. Once this requirement is satisfied, arbitrary conditions favourable to the simulations can be joined as much as necessary. In the present case, ordinary conditions of regularity for the retained potential and cancellation of the latter beyond the distance $r_C$, or cut-off distance, after which the effects of interaction potential are negligible, have been imposed.

The interaction potential is then defined by pieces on intervals $[0, r_0[$, $[r_0, r_C[$ and $[r_C, +\infty[$; it is continuous as is its first derivative at the meeting point of the attractive and repulsive parts (in $r_0$) and the attractive part and zero (in $r_C$). Its second derivative is also continuous in $r_0$ in order to cater for the eventuality of using this set to find the minima of the surface of potential energy by gradient methods. In normal use, distance $r_0$ and cut-off radius $r_C$ are fixed, so that this set of potentials comprises one parameter. It is denoted by $E_l(r)$; the parameter $l$ is defined at the end of this section.

The set of interaction potentials $E_l(r)$ was therefore constrained by the five conditions:

$$\begin{cases} E_l(r_0) = -\varepsilon \\ \dfrac{dE_l}{dr}(r_0) = 0 \\ \dfrac{d^2E_l}{dr^2}(r_0) = 72\dfrac{\varepsilon}{r_0^2} \\ E_l(r_C) = 0 \\ \dfrac{dE_l}{dr}(r_C) = 0 \end{cases} \quad (2)$$

The decision to use rational fractions to describe the potential attractive part was taken in order to have lightened computations while reproducing the sigmoid shape of the Lennard-Jones potential attractive part. Rational fractions, the number of coefficients of which is compatible with the number of conditions (2), were analysed and the choice of numerator and denominator degrees was decided according to the potential shape. This resulted in the set of interaction potentials of the following general mathematical form:

$$E_l(r) = \begin{cases} \varepsilon \cdot \left[ \left(\dfrac{r_0}{r}\right)^{12} - 2 \cdot \left(\dfrac{r_0}{r}\right)^{6} \right] & \text{for } 0 < r \leq r_0 \\ \varepsilon \cdot \dfrac{(r-A)^2 \cdot (r-B)}{C \cdot (r-r_0)^2 + D \cdot (r-r_0) + E} & \text{for } r_0 < r < r_C \\ 0 & \text{for } r_C \leq r \end{cases} \quad (3).$$

The expressions of constants $A$, $B$, $C$, $D$ and $E$ could be deduced from conditions (2). By posing $\rho = r_C/r_0 - 1$ and $\lambda = l/r_0 - 1$, they are written

$$\begin{cases} A = r_C \\ B = l \\ C = (36 \cdot \rho^2 \lambda + 2 \cdot \rho + \lambda) r_0 \\ D = -(\rho + 2\lambda)\rho \cdot r_0^2 \\ E = \rho^2 \lambda \cdot r_0^3 \end{cases} \quad (4).$$

It can be seen from expressions (3) that parameter $l$ may be taken as the root of equation $E_l(r) = 0$ other than $r_C$. It was possible to verify that the potential shape was similar to that of the Lennard-Jones potential for all values of parameter $l$ inferior to a fixed upper limit $l_{\max}$ dependent on $r_0$ and $r_C$.

**B. Numerical simulations**

These computer Molecular Dynamics simulations were performed using DLPOLY package (Ref.8), from systems exclusively composed of monoatomic particles having the characteristics of argon atoms. The particles evolved in a cubic simulation box with periodic boundary layer conditions; calculations were performed with time step $\Delta t_{cal} = 0.005\,\text{ps}$ and cut-off distance $r_C = 10.215\,\text{Å}$. Following the general hypotheses of CNT, the nucleations here considered are induced in systems maintained in isobaric and isothermic conditions, at temperatures strictly inferior to the melting temperature $T_m$. The starting time of the phenomenon is that at which the system reaches the required temperature conditions, after a rapid isobaric quench in the liquid state. The simulations are all carried out at $p = 420.6\,10^5\,\text{Pa}$. All quantities are expressed in the argon atom units system. The latter has $\sigma_{Ar} = 3.405\,\text{Å}$, $m_{Ar} = 6.642\,10^{-26}\,\text{kg}$, $\varepsilon_{Ar} = 1.6605\,10^{-21}\,\text{J}$ as primary quantities and includes the Boltzmann constant $k_B = 1.3807 \cdot 10^{-23}\,\text{J.K}^{-1}$. From these quantities, units of time $t_{Ar} = m_{Ar}^{1/2} \varepsilon_{Ar}^{-1/2} \sigma_{Ar} = 2.153\,\text{ps}$, pressure $p_{Ar} = \varepsilon_{Ar} \sigma_{Ar}^{-3} = 420.6\,10^5\,\text{Pa}$ and temperature $T_{Ar} = \varepsilon_{Ar} k_B^{-1} = 120.27\,\text{K}$ can be found. In the rest of this article, an upper asterisk denotes the value of the quantity expressed in this units system. As an example, we have $\Delta t_{cal}^* = 0.00232$, $r_C^* = 3$, $p^* = 1$ and $r_0^* = 1.122$. The implemented interaction potentials are the Lennard-Jones potential $E_{LJ}(r)$ given by (1) and the three potentials $E_{0.10}(r)$, $E_{0.95}(r)$ and $E_{1.05}(r)$ of the parameterized set defined by (3), where the lower index denotes $l^* = l/\sigma_{Ar}$ (Fig. 1).

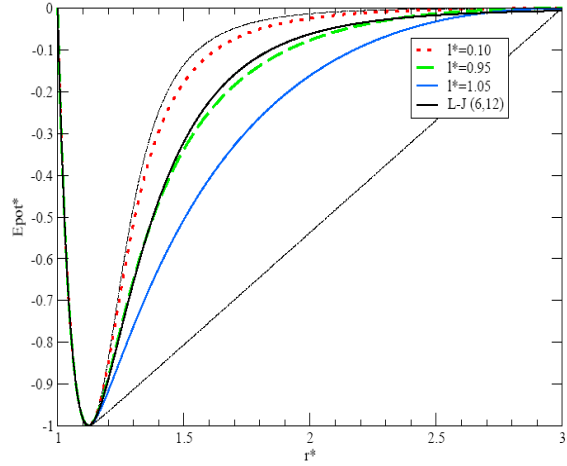

Figure 1. Implemented interaction potentials

Implemented interaction potentials $E_{0.10}(r)$, $E_{0.95}(r)$, $E_{1.05}(r)$ and $E_{LJ}(r)$ are situated in relation to the extreme potentials $E_{-10^9}(r)$ (left) and $E_{1.1038}(r)$ (right), where $l^*_{max}=1.1038$ when $r^*_C=3$; they are presented from left to right in order of increasing attraction.

It should be noted that the order relating to the increasing $l^*$ corresponds to the increasing attractions and that $l^*=0.95$ gives an interaction potential very close to the Lennard-Jones potential. In Ref. 7, these potentials are respectively denoted $E_I(r)$, $E_{II}(r)$ and $E_{III}(r)$ in the order of increasing attractions.

The numerical simulations of the present study are common with Ref. 7, except the average nucleation durations detailed below. For easy reading, implementation conditions and the results used in the following are briefly recalled. The thermal evolutions of diffusivity $D$ and particle densities of the crystal and liquid phases $\rho_C$ and $\rho_L$ stem from Molecular Dynamics numerical simulations of systems composed of 864 particles. They were obtained from cycles beginning with a fast heating from the crystalline state to the liquid state, followed by a fast quench from this liquid state to the starting crystalline state. Heating and quench consisted of successive isochronic stages of duration $\Delta t^*=464.5$, separated by increasing (heating) or decreasing (quench) temperature jumps $\Delta T^*=0.1$. The thermal correlations of the obtained $D^*$ values are given in Tab. I and those of $\rho^*_C$ and $\rho^*_L$ in Tab. II.

| potential | $E_{0.10}(r)$ | $E_{0.95}(r)$ | $E_{1.05}(r)$ | $E_{LJ}(r)$ |
|---|---|---|---|---|
| $a_D$ | - 3.0563 | - 2.9688 | - 3.0891 | -2.8942 |
| $b_D$ | 0.7248 | 0.3485 | 0.06571 | 0.2185 |

Table I. Coefficients of the thermal correlations of diffusivity by Arrhenius law
$$D^* = \exp(a_D/T^* + b_D)$$

| potential | $E_{0.10}(r)$ | $E_{0.95}(r)$ | $E_{1.05}(r)$ | $E_{LJ}(r)$ |
|---|---|---|---|---|
| $a_L$ | - 0.4135 | - 0.3308 | - 0.2825 | -0.3190 |
| $b_L$ | 1.1360 | 1.1228 | 1.1553 | 1.1177 |
| $a_C$ | - 0.1072 | - 0.09111 | - 0.08111 | -0.09417 |
| $b_C$ | - 0.05194 | - 0.09056 | - 0.08356 | -0.09072 |
| $c_C$ | 1.0487 | 1.0996 | 1.1539 | 1.1003 |

Table II. Coefficients of the thermal correlations of particle densities by formulas
$$\rho_L^* = a_L T^* + b_L \text{ and } \rho_C^* = a_C T^{*2} + b_C T^* + c_C$$

The thermal evolutions of Gibbs particle enthalpy of crystallization $\Delta G$, linked to Gibbs volumic enthalpy of crystallization $\Delta G_V$ by relation $\Delta G = \Delta G_V / \rho_C$, stem from the application of the $\lambda$-integration method of Grochola (Ref. 9) on systems composed of 864 particles, in four distinct temperatures and with the liquid and crystalline volumes resulting from the above simulations. The thermal correlations of the obtained $\Delta G^*$ values are given in Tab. III.

| potential | $E_{0.10}(r)$ | $E_{0.95}(r)$ | $E_{1.05}(r)$ | $E_{LJ}(r)$ |
|---|---|---|---|---|
| $a_G$ | - 0.8902 | - 0.3269 | - 0.1950 | -0.2017 |
| $b_G$ | - 0.4214 | - 1.0155 | - 1.1609 | -1.1480 |
| $c_G$ | 0.7841 | 0.9789 | 1.1807 | 1.0160 |

Table III. Coefficients for the interpolation of crystallization Gibbs particle enthalpies by
formula $\Delta G^* = a_G T^{*2} + b_G T^* + c_G$

The melting temperatures in Tab. IV come from the application of the coexisting phases method to a system composed of $10^6$ particles. Melting crystalline particle volumes are deduced from the thermal correlations in Tab. II. Results are in agreement with those in Ref. 10, accounting for the fact that calculations here were performed at $p^*=1$ and not -0.01.

| potential | $E_{0.10}(r)$ | $E_{0.95}(r)$ | $E_{1.05}(r)$ | $E_{LJ}(r)$ |
|---|---|---|---|---|
| $T_m^*$ | $0.77 \pm 0.03$ | $0.81 \pm 0.03$ | $0.89 \pm 0.03$ | $0.79 \pm 0.03$ |
| $V_{C,m}^*$ | 1.0489 | 1.0253 | 0.9839 | 1.0280 |

Table IV. Melting temperatures and particle crystal volumes according to implemented interaction potentials

Melting interfacial tensions in Tab. V result from the application of the capillary fluctuations method in Ref. 11 to a system composed of $10^6$ particles. The interface geometry was determined by the use of the local order parameter in Ref. 12 and the anisotropy corrections were obtained by means of a cubic limited expansion as in Ref. 13. The presented results correspond to the first term of the interfacial tension expansion. These results agree with those in Refs. 12 and 14.

| potential | $E_{0.10}(r)$ | $E_{0.95}(r)$ | $E_{1.05}(r)$ | $E_{LJ}(r)$ |
|---|---|---|---|---|
| $\gamma_m^*$ | 0.351 | 0.370 | 0.438 | 0.372 |

Table V. Interfacial tension according to interaction potential

The values of average nucleation durations $\langle t_{cx}^* \rangle_i$ stem from numerical simulations of systems composed of 108,000 particles at the studied temperatures $T_i^*$. They only made use of the interaction potentials $E_{0.10}(r)$, $E_{0.95}(r)$ and $E_{1.05}(r)$, simulations performed with $E_{LJ}(r)$ being identical to those of $E_{0.95}(r)$. Unlike the method in Ref. 7, all average durations $\langle t_{cx}^* \rangle$ at which the first crystalline nucleus appears in the simulation volume were deduced from durations $t_{cx}^*$, at which the first drop in the system's potential energy was recorded. Computation of $\langle t_{cx}^* \rangle_i$ was performed when the 10 values of $t_{cxi}^*$ related to $T_i^*$ were available. Fig. 2 shows the obtained results.

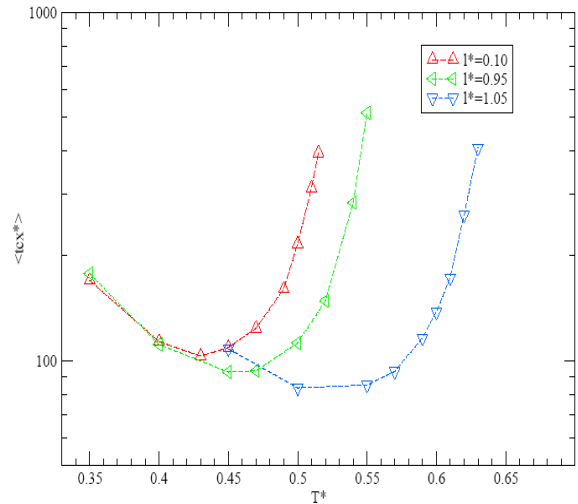

Figure 2. Simulated average nucleation durations

Simulation results as symbols and dotted lines for easy reading

The thermal evolutions at high temperatures coincide with those in Ref. 7, while they are slightly lower at low temperatures. The global shape of the evolutions is similar to that found in literature, in particular in Refs. 15 and 16.

**C. Thermal evolutions of interfacial tension**

The present method to determine the thermal evolutions of interfacial tensions related to the tested attractions will require, in an essential way, the formalism of CNT. Therefore, in the hypothesis made here of an homogeneous nucleation regarded as the random appearance of the nuclei number $dN$ in the volume $dV$ of the system during the duration $dt$ at time $t$, the instantaneous nucleation rate $J(t)$ defined by $J(t) = 1/V \cdot dN/dt$ constitutes the obliged starting point of the reasoning. In other respects, experiments show that temporal evolutions of crystallization nuclei densities in isobaric and isothermal conditions progress in three successive stages: incubation, free nucleation and saturation (Ref. 17). The Zeldovich formula (Ref. 4)

$$J(t) = J_0 \cdot e^{-\frac{\tau}{t}} \qquad (5),$$

where $\tau$ is the time constant and $J_0$ the stationary nucleation rate, appear then as the ideal tool for an unsteady-state global approach of nucleation before saturation, the exponential factor perfectly reproducing the succession of incubation, namely a null nucleation rate on a finite duration, and free nucleation. The relevant size for the pursuit of the calculation being the number of nuclei $N$ which appeared in the studied volume $V$ at time $t$, integration from the definition according to $t$ and reporting $J(t)$ given by (5) lead to :

$$N = J_0 V t \cdot Ei_2\left(\frac{\tau}{t}\right) \qquad (6),$$

where $Ei_2(x) = \int_1^{+\infty} y^{-2} e^{-xy} dy$ is the special function integral exponential of order 2 (Ref. 18). The application of this formula to our numerical simulations, where the nucleation duration $t_{cx}^*$ corresponds to the appearance of $N=1$ nucleus in the simulation volume $V$ and time $t$ to the average $\langle t_{cx}^* \rangle$ of $t_{cx}^*$, gives

$$1 = J_0^* V^* \langle t_{cx}^* \rangle \cdot Ei_2\left(\frac{\tau^*}{\langle t_{cx}^* \rangle}\right) \qquad (7).$$

Among the significant contributions of CNT figure the expressions deduced from Ref. 17

$$\tau = \frac{2^{\frac{1}{3}}}{3}\left(\frac{2\pi}{3}\right)^{\frac{4}{3}} \cdot \frac{\rho_C^{\frac{4}{3}} \gamma^4}{\rho_L^{\frac{2}{3}} \Delta G_V^4 D} \qquad (8)$$

and

$$J_0 = \left(\frac{3072}{\pi}\right)^{\frac{1}{3}} \cdot \frac{\rho_L^{\frac{5}{3}} \gamma^{\frac{1}{2}} D}{\rho_C^{\frac{1}{3}} (k_B T)^{\frac{1}{2}}} e^{-\frac{16\pi}{3} \cdot \frac{\gamma^3}{\Delta G_V^2 k_B T}} \qquad (9)$$

of the parameters of the Zeldovich formula. By reporting in (7) and knowing that $V^* = N_{part}/\rho_L^*$, the relation

$$1 = \left(\frac{3072}{\pi}\right)^{\frac{1}{3}} \frac{\rho_{L,i}^{*\frac{2}{3}} \gamma_i^{*\frac{1}{2}} D_i^*}{\rho_{C,i}^{*\frac{1}{3}} T_i^{*\frac{1}{2}}} e^{-\frac{16\pi}{3} \cdot \frac{\gamma_i^{*3}}{\Delta G_{V,i}^{*2} T_i^*}} N_{part} \langle t_{cx}^* \rangle_i Ei_2\left(\frac{2^{\frac{1}{3}}}{3}\left(\frac{2\pi}{3}\right)^{\frac{4}{3}} \frac{\rho_{C,i}^{*\frac{4}{3}} \gamma_i^{*4}}{\rho_{L,i}^{*\frac{2}{3}} \Delta G_{V,i}^{*4} D_i^* \langle t_{cx}^* \rangle_i}\right) \qquad (10)$$

can be deduced, allowing the determination of the interfacial tension thermal evolutions.

Indeed, it can be remarked that CNT reduces the solving of a nucleation problem to the data of interfacial tension $\gamma$, crystallization Gibbs volumic enthalpy $\Delta G_V$, diffusivity $D$, particle densities $\rho_L$ and $\rho_C$ of the liquid and crystalline phases and nucleus thermal fluctuations energy $k_B T$. The thermal correlations recalled in Sect. B allowing to determine, for each tested temperature $T_i^*$, the corresponding values $D_i^*$, $\Delta G_{V,i}^*$, $\rho_{L,i}^*$ and $\rho_{C,i}^*$ and the values $\langle t_{cx}^* \rangle_i$ stemming from the nucleation simulations, interfacial tension $\gamma_i^*$ related to $T_i^*$ figures as the only unknown of (10). Solved in $\gamma_i^*$, this equation provided, for each implemented interaction potential, the dimensionless sizes $\hat{\gamma} = \gamma^*/\gamma_m^*$ and $\hat{T} = T^*/T_m^*$, represented in Fig. 3.

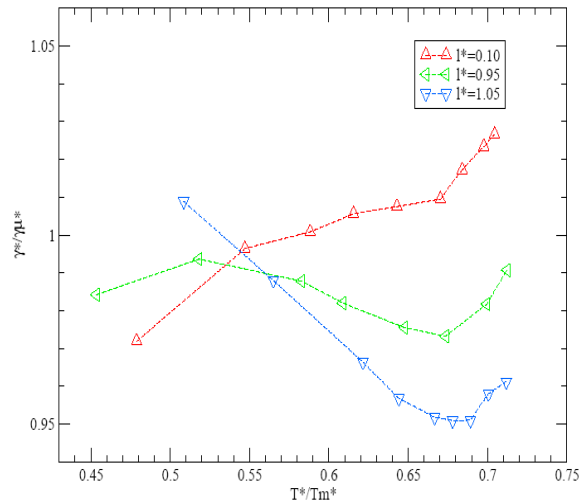

EMBED AcroExch.Document.7

Figure 3. Interfacial tension consistent with the CNT
Simulation results as symbols and dotted lines for easy reading.

**D. First remarks and comments**

First, it should be remarked that, in view of the computation accuracy and regularity of their representative curve, the thermal evolutions presented in Fig. 3 have nothing erratic. So, they are significant and consequently perfectly reflect CNT.

A remark stands out concerning the range of the three thermal evolutions: overall the interval of the explored temperatures (around 50K), the reduced interfacial tension variations occur in a limited domain, comprised between 0.95 and 1.05. Knowing that all these evolutions move through the point of coordinates $(T/T_m = 1, \gamma/\gamma_m = 1)$, one can consider that the law which expresses best the thermal evolution of interfacial tension in coherence with CNT is the constant law $\gamma(T) = \gamma_m$. But this law is only valid to within $\pm 0.05$, so as in formulas (7) or (8) or (9), it can provide, because of the exponential and integral exponential functions, nucleation rate values to within several orders of magnitude.

Looking at these evolutions more finely, it can be observed, as was attempted by considering the evolutions of average nucleation durations in Fig. 2, and those of diffusivity, particle densities of liquid and crystal phases and crystallization Gibbs enthalpy provided in Ref. 7, that the thermal evolutions presented in Fig. 3 are obviously sensitive to the effects of potential attraction. Nothing more is necessary to make the thermal evolution of interfacial tension a parameter of the theory. But, for all that, can it be considered as a fully physical size?

Contrary to what is observed for all sizes from which it is deduced (see Fig. 3 and Ref. 7), interfacial tension does not behave in a regular manner: according to the attraction value, it can increase monotonously ($l^*$=0.10); increase slightly, then decrease before increasing again ($l^*$=0.95); or else globally decrease, then increase ($l^*$=1.05). Now, it is difficult to conceive that temperature and attraction intensity could have such unclassifiable effects on a quantity that is simply geometrically linked to the forces of attraction. One cannot but conclude that interfacial tension consistent with CNT does not have the same physical nature as true physical sizes such as diffusivity, particle density or crystallization Gibbs enthalpy.

It is now possible to answer the question of the status of the interfacial tension in CNT. If forecasts within several orders of magnitude can be accepted, interfacial tension is a fully physical size and this coincides with the melting interfacial tension within $\pm 0.05$ in relative values. If more accurate forecasts are needed, then, and because of the unclassifiability of their thermal behaviours, interfacial tension cannot be considered as a true physical size, but, at the most, as a parameter the thermal evolutions of which are to be modelled.

At the end of these remarks and comments, the conclusion that seems to stand out is that, as the above comments finally concern CNT (the thermal laws of interfacial tension are, in fact, its reflection), this theory must conceal a flaw which definitively invalidates the accuracy of its forecasts. Now, this theory possesses great qualities, the most notable of which are the relevance of its physical analysis, its formal simplicity and its efficiency. If it were to be replaced, these qualities should be maintained.

## III. A REPLACEMENT SOLUTION FROM THE ZELDOVICH FORMULA

### E. Problem setting without dimension anomaly

We retain as first base for this new nucleation problem setting, the analysis from CNT following which the variables dimensionally characteristic of nucleation are $\gamma$, $\Delta G_V$, $D$, $k_B T$ and $\rho_L$. The second base is the Zeldovich formula under form (5) for the purely formal reason that it combines simplicity and reliability.

In the International Units System, the above nucleation variables are expressed respectively as J.m$^{-2}$, J.m$^{-3}$, m$^2$.s$^{-1}$, J.C$^{-1}$ and P.m$^{-3}$. The respective units C and P for the number of nuclei (clusters) and particles have had to be introduced here in order to account for the specificity of the nucleation phenomenon. Indeed, in a question of nucleation, a nucleus cannot be assimilated to a particle, and conversely, so these numbers are two distinct quantities and consequently, must be measured in two distinct units, C and P, as proposed here. Thus, it can be noted that the factor $N = -(16\pi/3)\cdot\gamma^3/(\Delta G_V^{\,2} k_B T)$ under the exponential in (9) does indeed have the dimension C of a nuclei number, but $\tau$ is in $P^{\frac{2}{3}}.s$, rather than in s as $t$ and the factor $M = \left(\rho_L^{\frac{5}{3}}\gamma^{\frac{1}{2}}D\right)\Big/\left(\rho_C^{\frac{1}{3}}(k_B T)^{\frac{1}{2}}\right)$ of $J_0$ is in $C^{\frac{1}{2}}.P^{\frac{4}{3}}.m^{-3}.s^{-1}$ rather than in C.m$^{-3}$.s$^{-1}$ as $J(t)$. In this new nucleation problem setting, the choice is made to resolve this anomaly. The question is then to know if there is a combination of the nucleation variables leading to the correct units of $\tau$ and $J_0$.

The units system characteristic of nucleation is $([\gamma],[\Delta G_V],[D],[k_BT],[\rho_L])$ where the brackets denote the unit of the physical size. According to the above, this set of units has $(m,s,J,C,P)$ as its fundamental units system. Now, Vaschy-Buckingham's theorem shows that there is only one way to express the second units according to the first units, the conversion matrix from the units system characteristic of nucleation to the fundamental units system

$$\begin{pmatrix} -2 & -3 & 2 & 0 & -3 \\ 0 & 0 & -1 & 0 & 0 \\ 1 & 1 & 0 & 1 & 0 \\ 0 & 0 & 0 & -1 & 0 \\ 0 & 0 & 0 & 0 & 1 \end{pmatrix}$$

being of rank 5. The resolution leading to

$$\begin{cases} m = [\gamma][\Delta G_V]^{-1} \\ s = [\gamma]^2.[\Delta G_V]^{-2}.[D]^{-1} \\ J = [\gamma]^3.[\Delta G_V]^{-2} \\ C = [\gamma]^3.[\Delta G_V]^{-2}.[k_BT]^{-1} \\ P = [\gamma]^3.[\Delta G_V]^{-3}.[\rho_L] \end{cases}.$$

and the units now imposed on $\tau$, $M$ and $N$ being respectively $s$, $C.m^{-3}.s^{-1}$ and $C$, the only acceptable groups of characteristic quantities of nucleation are respectively $\gamma^2/(\Delta G_V^2 D)$, $D\Delta G_V^3/(\gamma^2 k_BT)$ and $\gamma^3/(\Delta G_V^2 k_BT)$. The last group is identical to that occurring in formula (9) derived from the CNT. It is therefore relevant to suppose that $N = -(16\pi/3).\gamma^3/(\Delta G_V^2 k_BT)$. Concerning $\tau$ and $J_0$, it can only be written that

$$\tau = L.\frac{\gamma^2}{\Delta G_V^2 D} \tag{11},$$

and

$$J_0 = K.\frac{D\Delta G_V^3}{\gamma^2 k_BT}.e^{-\frac{16\pi}{3}.\frac{\gamma^3}{\Delta G_V^2 k_BT}} \tag{12},$$

where $L$ and $K$ are two constants unattainable by this reasoning of dimensional analysis. At this stage of the new problem setting, expressions (11) and (12) may be integrated in formula (7) adapted to our numerical simulations. With the values of the physical sizes resulting from the numerical simulations and expressed in the units system of argon atom, one obtains

$$1 = K \frac{D_i^* \Delta G_{V,i}^{*3}}{\gamma_i^{*2} T_i^*} e^{-\frac{16\pi}{3} \frac{\gamma_i^{*3}}{\Delta G_{V,i}^{*2} T_i^*}} \frac{N_{part}}{\rho_{L,i}^*} \langle t_{cx}^* \rangle_i Ei_2 \left( L \frac{\gamma_i^{*2}}{\Delta G_{V,i}^{*2} D_i^* \langle t_{cx}^* \rangle_i} \right) \tag{13}$$

as relation corresponding to (10).

Both constants $K$ and $L$ must obviously be supposed to be universal, that is to say, in the precise case of the present simulations, to not depend on interaction potential attraction and on temperature. It is to be seen that, $K$ and $L$ being unknowns, relation (13) does not enable us, as in Sect. C, to determine the values of $\gamma_i^*$ on the basis of data from numerical simulations and thermal correlations from Sect. B. The only way to resolve this indetermination is to dictate the mathematical form of the law $\gamma(T)$.

**F. Physical hypotheses and mathematical formulation of a simple and realistic thermal evolution law of interfacial tension**

The question of a mathematical form for modelling the thermal evolution of interfacial tension is here posed. In order to reduce the arbitrary nature of such a choice, a few general physical hypotheses will now be ventured. We first assume, in agreement with CNT (Ref. 19), that the interfacial tension $\gamma$ to which nucleation phenomena correlate is the interfacial tension between the liquid parent phase and the crystalline wall of nuclei. This is in accordance with the calculations below, using Formulas (11) and (12) for which $\gamma$ must only be in $J.m^{-2}$. Thus, we assume that the thermal evolution of this quantity is the same as the other characteristic quantities of nucleation (Ref. 7), in that it does not undergo any singularity when crossing the melting point (hypothesis n°1). This only assumes the possibility of overheated metastable crystalline nuclei in the liquid in the same way that crystalline nuclei can be observed in the super-cooled metastable liquid. As the thermal evolution of $\gamma$ must thus be monotonous and the attractive centres, in sliding apart as temperature increases, produce a decrease in the attractive forces, this evolution will be assumed, in the same way as the tension at the liquid-vapour interface (Refs. 20 and 21), to decrease with temperature in the domain of super-cooled liquids and be almost linear in temperature in the super-cooled liquid domain near melting (hypothesis n°2). It can be deduced that interfacial tension reaches its maximum value at the limit of absolute null temperature. These first two hypotheses in accordance with the approach of the geometrical nature of the interfacial tensions mentioned in the introduction, lead to think that interfacial tension at absolute null temperature is all the greater as the attraction between particles is high (hypothesis n°3), and that the slope of the thermal evolution of interfacial tension at melting point vicinity is all the lower since the attraction is high (hypothesis n°4). Finally, as classical laws describing the behaviour of matter are generally scale invariant, such an assumption for interfacial tension should be roughly appropriate, as the criterion of spatial correlation to infinity (Ref. 22) is satisfied in the case of a liquid with or without nuclei, the particles being small in comparison with the Van der Waals forces acting radius. However, the case of absolute null temperature must be examined. The phonons by which energy spreads inside the particle lattice being discrete, a number of clusters of moving particles in a growing motionless set should be observed as temperature decreases. In this situation, the criterion of spatial correlation to infinity becomes clearly incorrect. Because of this limit-case, the interfacial tension law on the overall domain of practicable temperatures should be scale dependant. In the following, we assume that this is the case (hypothesis n°5). These hypotheses being stated, the general mathematical form of the thermal evolution of interfacial tension remains to be deduced.

Since absolute zero in temperature has a central role in the framework of hypothesis n°5, the variables $\gamma_0 - \gamma$ and $T$, centred on interfacial tension $\gamma_0 = \gamma(0)$ and $T = 0$ K, are to be considered. The physical law being described with the wider generality from $(\gamma_0 - \gamma)/\gamma_m$ and $T/T_m$, which are the forms of $\gamma_0 - \gamma$ and $T$ reduced to melting conditions, the general form of the sought-for law $\gamma(T)$ is written as $\hat{\gamma} = \hat{\gamma}_0 - f(\hat{T})$, denoting $\hat{T} = T/T_m$ and $\hat{\gamma} = \gamma/\gamma_m$. With these notations and to prepare the application of hypothesis n°5, the mathematical form $\hat{\gamma} = \hat{\gamma}_0 - b.\hat{T}^{k+1}$ of a scale invariant law is recalled (Ref. 22). In other respects, the simplified linear local form, as in Refs. 20 and 21 for liquid-vapour interfacial tension, here written as

$$\hat{\gamma} = 1 - b.(1 - \hat{T}) \qquad (14)$$

to express the fact that $\hat{\gamma}(1) = 1$ (*cf.* hypothesis n°1), will be used. To obtain the mathematical expression of an observation scales dependant law (Ref. 23), we remark that the variables $\hat{\gamma}_0 - \hat{\gamma}$ and $\hat{T}$ of the sought-for law are essentially differences, and that the values $\hat{\gamma}$ and $\hat{T}$ may only be reached when taking the scale effects into account, by relating these finite differences to the infinitesimal corresponding differences $d\hat{\gamma}$ and $d\hat{T}$. Clearly having the proportionality relation $d\hat{\gamma} = m.d\hat{T}$, where $m$ is a constant, the rate $d\hat{\gamma}/(\hat{\gamma}_0 - \hat{\gamma})$, which can also be written $m.d\hat{T}/(b\hat{T}^{k+1})$, provides $d\hat{\gamma}/(\hat{\gamma}_0 - \hat{\gamma}) = m.d\hat{T}/(b\hat{T}^{k+1})$. Through integration, this differential equation leads to the searched-for relation between the quantities $\hat{\gamma}$ and $\hat{T}$. Thus, provided that $k > 1$, a scale dependant law has the form of a stretched Arrhenius law $\hat{\gamma} = \hat{\gamma}_0 - A\exp(-(B/\hat{T})^k)$, where $A$ is the integration constant and $B = (m/(b(k-1)))^{\frac{1}{k-1}}$. In the present case of an evolution crossing the melting point (*cf.* hypothesis n°1), $\hat{\gamma}(1) = 1$ must also be verified. This being done when $A = (\hat{\gamma}_0 - 1)e^{B^k}$, the thermal evolution law of interfacial tension consistent with hypotheses n°5 and 1 is finally as

$$\hat{\gamma} = \hat{\gamma}_0 - (\hat{\gamma}_0 - 1)\exp(B^k.(1 - 1/\hat{T}^k)) \qquad (15).$$

### G. The proposed interfacial tension law

Taking up again the problem as posed at the end of Sect. E and taking account of interfacial tension laws (14) et (15), the question now becomes to specify the thermal interfacial tension law, namely to determine universal constants $K$ and $L$, and the three adjusting parameters $\hat{\gamma}_{0,l^*}$, $B_{l^*}$ and $k_{l^*}$ related to each of the three attractions $l^*$=0.10, 0.95, 1.05, which verify equation (13). This problem could be specified in the following way: to find, on the basis of a set of pairs $(T_i^*, \langle t_{cx}^* \rangle_i)$ from the numerical experiments and thermal correlations giving $D_i^*$, $\Delta G_{V,i}^*$, $\rho_{L,i}^*$ et $\rho_{C,i}^*$ (Sect. B), the set of 11 adjusting parameters $K$, $L$, $\hat{\gamma}_{0,l^*}$, $B_{l^*}$ and $k_{l^*}$ that minimizes the average difference $\Delta$ between on the one hand, the interfacial tension values $\hat{\gamma}_i^*$ achieved from (13) for this set of parameters and, on the other hand, those of $\hat{\gamma}(\hat{T}_i)$ from its modelling for the same set of parameters by one of the correlations (14) or (15), whether an approximate or exact resolution is desired; $\Delta$ is precisely defined by

$$\Delta = \sum |\hat{\gamma}_i - \hat{\gamma}(\hat{T}_i)|/N \qquad (16),$$

where $N$ is the number of value pairs $(T_i^*, \langle t_{cx}^* \rangle_i)$.

Preliminary numerical tests had underlined the existence of several local minima in $\Delta$, for which the values of adjusting parameters led to evolutions $\hat{\gamma}(\hat{T})$ in disagreement with one or several of our hypotheses. To force the search for the adjusting parameters onto a solution verifying hypotheses n°1 and 2, $\hat{\gamma}(\hat{T})$ was first modelled by the linear correlation (14) simulating the linear thermal evolution of the liquid in the super-cooled liquid domain near melting, the simulations retained to establish the values of constants $K$ and $L$ being the nearest at the melting point. Thus, the linear model (14) validated the hypotheses n°1, 2 and 4, but invalidated hypothesis n° 3. By modelling the thermal evolutions by law (15), hypothesis n° 3 could be validated, while hypotheses n°1 and 5 were automatically validated by using (15) and finally, by retaining the obtained values of $K$ and $L$, hypotheses n°2 and 4 were also validated.

The search for optimization parameters was finally carried out using the following method: 1) setting the constants $K$ and $L$ at arbitrary values; 2) for each pair $\left(T_i^*, \langle t_{cx}^* \rangle_i\right)$ from the numerical experiments, computing the solution $\gamma_i^*$ of equation (13) completed by the thermal correlations related to $D_i^*$, $\Delta G_{V,i}^*$, $\rho_{L,i}^*$ et $\rho_{C,i}^*$ (Sect. B); 3) from those values of $\gamma_i^*$, computing the value of adjusting parameters $b_{l^*}$ by means of least mean square linear regressions, for each of the searched-for $l^*$, on the basis of approximated modelling by (14); this computation is performed by omitting the points at the four lowest temperatures; 4) testing several values of $K$ and $L$, the values finally retained being those that minimize the average difference $\Delta_{(14)}$ (formula (16)); 5) on the basis of those $K$ and $L$ values, and for each $l^*$ value, computing by least mean square linear regressions, the adjusting parameter $B_{l^*}$ appearing in (15) and corresponding to an arbitrarily fixed pair $\left(\hat{\gamma}_{0,l^*}, k_{l^*}\right)$; this computation was performed on the basis of each $\left(T_i^*, \langle t_{cx}^* \rangle_i\right)$ from the numerical experiments pairs; 6) testing several pairs $\left(\hat{\gamma}_{0,l^*}, k_{l^*}\right)$, the values finally retained being those that minimize the average difference $\Delta_{(15)}$ (formula (16)).

The above method led to $K = 0.896 \cdot 10^{15} \pm 5 \cdot 10^{11}$ and $L = 0.997 \cdot 10^{-1} \pm 5 \cdot 10^{-5}$ providing the minimal average global difference $\Delta = 0.00805$ on the basis of law (15). Details of the results are presented in Tab. VI. This provides, for each $l^*$ implemented, the values of the adjusting parameters and corresponding average difference $\Delta$.

| $l^*$ | $b$ | $\Delta_{(14)}$ | $\hat{\gamma}_0$ | $B$ | $k$ | $\Delta_{(15)}$ |
|---|---|---|---|---|---|---|
| 0.10 | 1.72848 | 0.01203 | 1.747 | 0.83323 | 3.288 | 0.00508 |
| 0.95 | 1.59094 | 0.00367 | 1.780 | 0.80969 | 2.750 | 0.00191 |
| 1.05 | 1.47457 | 0.00864 | 1.790 | 0.79870 | 2.450 | 0.00265 |

Table VI. Adjusting parameters according to interaction potential attraction with the values $K = 0.896 \cdot 10^{15} \pm 5 \cdot 10^{11}$ and $L = 0.997 \cdot 10^{-1} \pm 5 \cdot 10^{-5}$ minimizing $\Delta$ to 0.00805

Fig. 4-a) represents the interfacial tension calculated from the simulations by relation (13) and thermal correlations recalled in Sect. B, $K$ and $L$ having the above values. In Fig. 4-b) to 4-d) for a fixed $l^*$, the above points from the simulations and evolutions modelled by (15) on the basis of the values of the adjusting parameters presented in Tab. VI, are superimposed, with the above values as universal ones for $K$ and $L$.

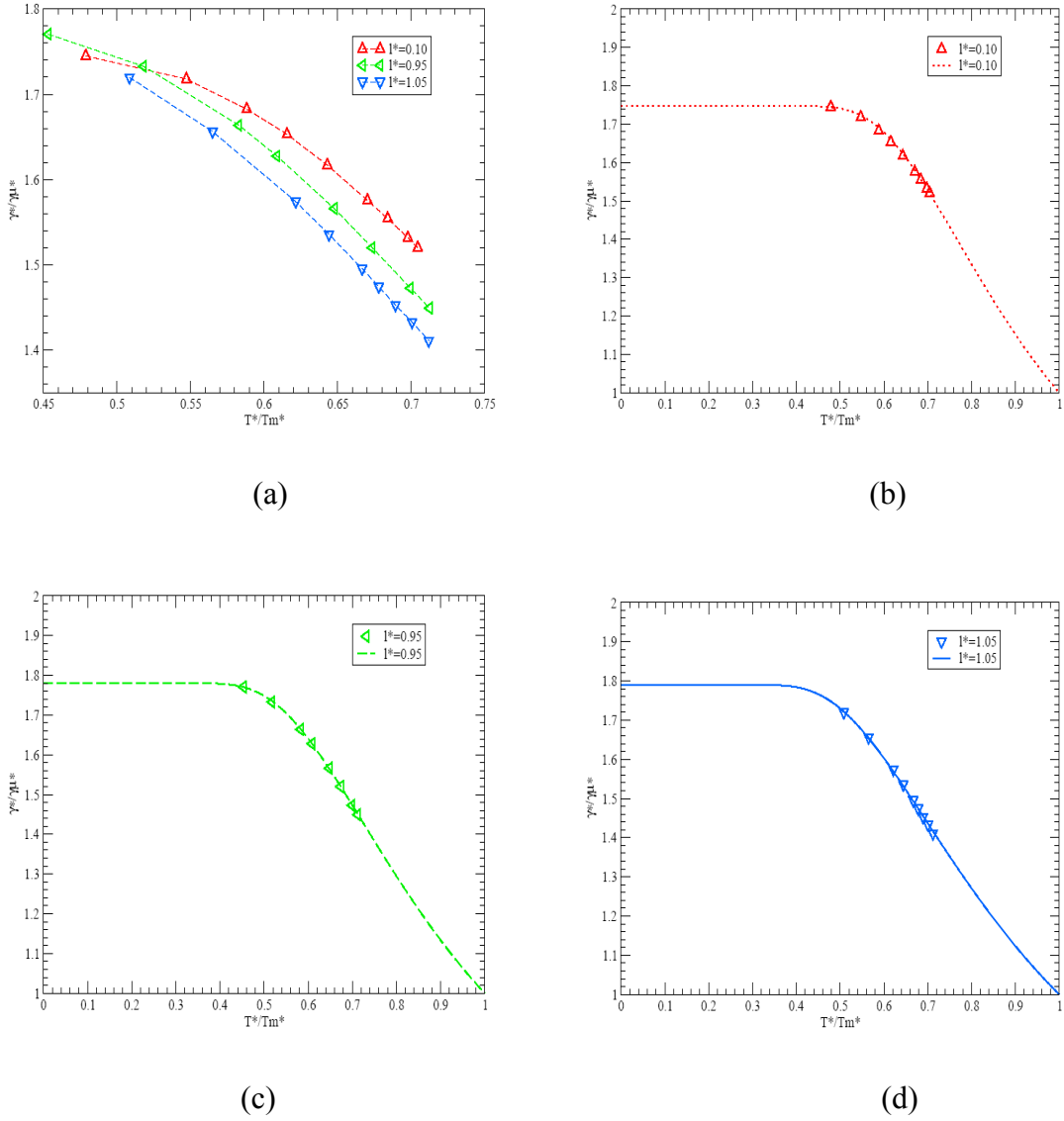

Figure 4. Results in non-dimensional sizes
Simulation results as symbols; (a): dotted lines for easy reading, (b), (c) and (d): law as solid line

The adjusting parameters $\hat{\gamma}_{0,l^*}$, $B_{l^*}$ and $k_{l^*}$ of law (15) clearly depend on the attractivity of the interaction potential (Tab. VI). We assumed that the attractivity can be measured by the difference between an arbitrary test particle volume of the system and the volume $V_0$ it should have at vanishing potential attraction. Such an assumption can be justified by the analogy between the potential energy resulting from attraction forces and the work of a negative pressure acting from inside the system. By taking the melting crystal particle volume $V_{C,m}$ as the test volume, the attractivity can thus be defined here by the value $V_0^* - V_{C,m}^*$ of $V_0 - V_{C,m}$.

On the basis of the values in Tab. IV and VI and by using the least mean square linear regression method, the following correlations were reached

$$\begin{cases} \hat{\gamma}_0 = 1.79872 \cdot e^{2.60623 \cdot 10^{-4} \cdot \ln^3(1.057 - V^*_{C,m})} \\ B = 0.784906 \cdot e^{2.58066 \cdot 10^{-3} \cdot \ln^2(1.057 - V^*_{C,m})} \\ k = 1.73068(1.057 - V^*_{C,m})^{-0.133455} \end{cases} \qquad (17).$$

They are represented in Fig. 5. A correlation such as $k \propto (V_0^* - V^*_{C,m})^n$ was first investigated. The difference $(1.057 - V^*_{C,m})$ obtained was then retained as the variable for the other two correlations, with the constant 1.057 being interpreted as the value $V_0^*$ of the particle volume that the melting crystal should have at vanishing interaction potential attraction.

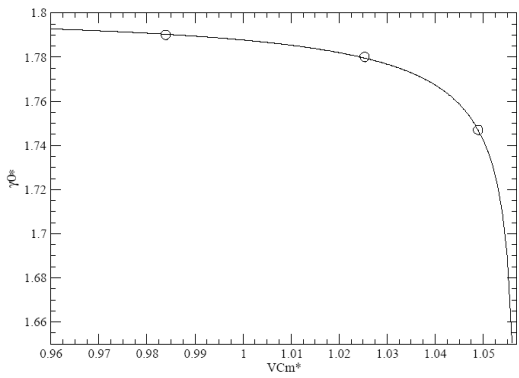

(a)

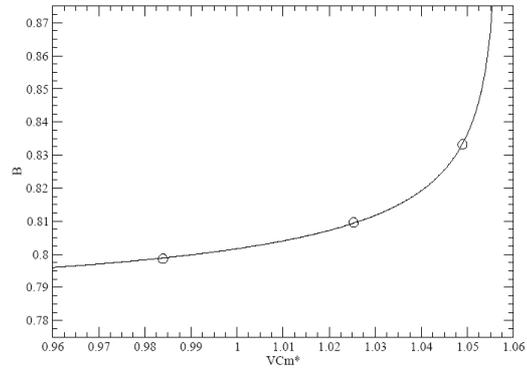

(b)

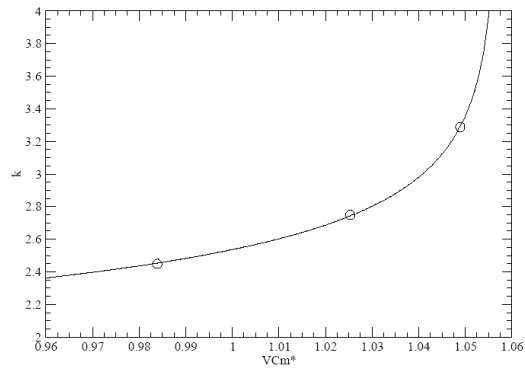

(c)

Figure 5. Correlations of the adjusting parameters of law (15) with the particle volume of melting crystal

As a final verification, Fig. 6 superimposes the average nucleation durations stemmed from simulations and their evolution obtained by the law (15)-correlations (17) set.

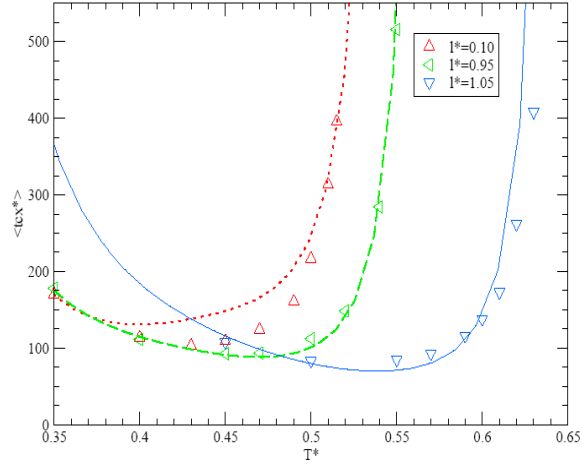

Figure 6. Accuracy of the forecasts of the interfacial tension law
Simulation results as symbols and forecasts using (15) and (17) as lines

## H. Second remarks and comments

The quality of the mathematical form of law (15) can be perceived by its ability to be near the points from simulations. Fig. 4-b) to 4-d) show a difference hardly discernible between the simulation points and their representative curve, the value of the global average difference $\Delta$ defined by (16) being 0.008. Since law (15) can finally be accurately adjusted with only the three parameters $\hat{\gamma}_0$, $B$ and $k$ to the three different attractivities, the adopted mathematical form seems *a posteriori* relevant. Concerning the quality of correlations (17), which can be assessed from Fig. 5-a) to 5-c), three adjusting parameters were used to be the nearest to three points, but only with two degrees of freedom, $V_0^*$ having the fixed value 1.057. The good accuracy leads to think that the choice of attractivity $(1.057 - V_{C,m}^*)$ to measure the attraction of interaction potential is relevant. Last, the quality of the interfacial tension law (law (15) and correlations (17)) may finally be assessed by the difference between forecasted values of average nucleation duration and values from the numerical experiments. Fig. 6 shows that relative error is in the order of a few percent at $l^*=0.95$, it rises to around 15% at $l^*=1.05$ and attains 30% for $l^*=0.10$. These results are to be compared with the forecasts within several orders of magnitude that can be found in the literature.

The comparison of the bundles of three thermal interfacial tension evolutions presented in Fig. 3 and 4-a) indicates that the alternative theory does not present the defect of CNT mentioned in Sect. D. Indeed, the thermal interfacial tension evolutions do no longer have their unclassifiable character according to attractivity (Sect. D), but indicate on the contrary a unity of behaviour. In this alternative theory, interfacial tension has become a physical size of the same nature as the other characteristic nucleation sizes. Moreover, in addition to a good coincidence with the simulation points, the thermal law in which Sect. G results agrees with the hypotheses stated in Sect. F: it can be observed that the part of this thermal law near melting has a linear shape as required by hypothesis n°2, the slope at melting point is, as required by hypothesis n°4, all the greater as attractivity is small and, at absolute zero in temperature, the interfacial tension all the higher as attractivity is high, following hypothesis

n°3. The ordinate $\hat{\gamma}_0$ =1.78 of the case $l^*$=0.95 similar to the Lennard-Jones case corresponds to the value $\gamma_0$ =0.66, roughly in agreement with the value 0.44 of the interfacial tension obtained by the numerical unstable sphere method at low temperature on metallic particles, by Morris and Song in Ref.12. Concerning hypothesis n°2, our search for parameters minimizing the difference $\Delta$ between simulation points and law (15) led to an alternative adjusting parameters set with which the thermal evolution law of interfacial tension was increasing, $\Delta$ having almost the same value, but invalidating hypotheses n°3 and 4. So, the fact that the thermal evolution of interfacial tension is here decreasing results in the choice to satisfy hypothesis n°2, which led to the validity of hypotheses n°3 and 4 as emerging properties of the model. If the physical character of the latter can be accepted, one sees that it is the same for law (15), and the fact that hypotheses n°2, 3 and 4 are so bounded gives a particular importance to the proposed alternative theory. Certainly, the growth of this law is contrary to that presented by several authors, among which Turnbull in Ref. 15, Peng *et al.* in Ref. 16 or Aga in Ref. 24, who found their reasoning onto the entropic free energy of crystal-liquid interface, but certain authors (Ref. 14) have proposed models with decreasing interfacial tension. To finish, let us notice that scales dependence which appears at very low temperatures expresses here by a long quasi-plateau.

## IV. IMPLEMENTATIONS

### J. Average nucleation durations

The implementation in equation (13) of the interfacial tension law (15) with thermal correlations of Sect. B enabled $\langle t^*_{cx} \rangle$ to be solved for each value of $T^*$ and the thermal evolution forecasts of $\langle t^*_{cx} \rangle$ to be established. Fig. 7 presents these thermal evolutions with durations of up to approximately one month, for the three implemented interaction potentials. The three curves have similar, basin-like shapes. Their rather flat lowest part is comprised between two vertical asymptotes, occurring clearly in $T^*$=0 and $T^* = T^*_{m,l^*}$. The effects of potential attraction are to slightly lower the minima and shift the ascending parts of the curves towards the right as the attraction increases, following the shift in the melting temperature (Tab. IV). Conversely, it has almost no effect on the descending part of the curves. The combined effect of these shifts and the very high slope leads, other than at the basin bottom, to the attraction influencing the average nucleation duration, measured in decades.

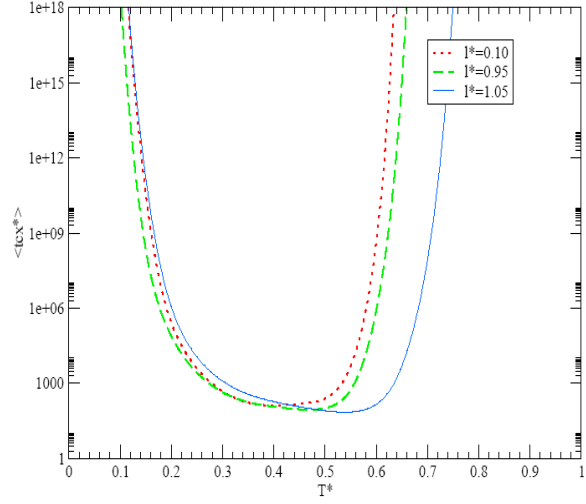

Figure 7. Forecasted evolutions of average nucleation durations

**K. Ratios of the stationary nucleation rate realized at nucleation**

Taking the average nucleation duration $\langle t_{cx}^* \rangle$ as in Sect. J and employing it in relation (5), after using formula (11), allowed us to express the fraction $J(\langle t_{cx}^* \rangle)/J_0$ simply, representing the rate at which nucleation takes place. Indeed, values close to zero correspond to nucleation occurring during incubation, *i.e.* in the transient rate, and those close to one correspond to the stationary rate. Fig. 8 represents the evolution of this fraction for the three implemented attractions. It shows that the stationary rate is realized in the extremes of the temperature domain, at the low and deep super-cooling liquids. The hollow shape is narrower than for average nucleation durations (Fig. 7). The question of the rate at which numerical nucleations occur (0.3< $T^*$ <0.6) is easily answered by means of Fig. 8. Indeed, this figure shows that the nucleations always occur in the transient rate, in accordance with the conclusion of Peng *et al.* in Ref. 16.

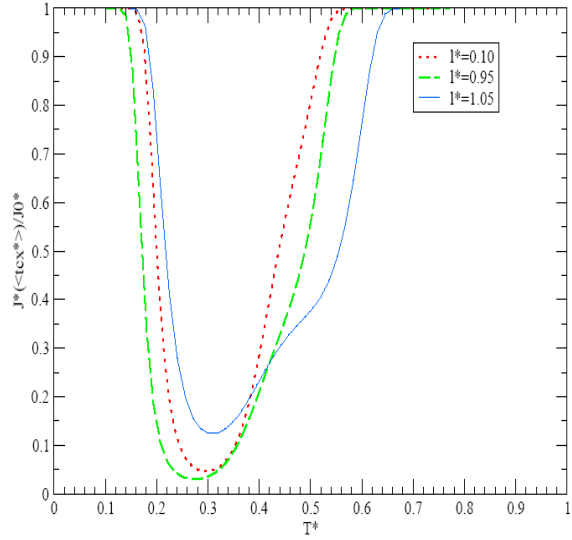

Figure 8. Forecasted evolutions of the fraction of the stationary nucleation rate attained at nucleation

**L. Realized nucleation rates**

Disposing of $\langle t_{cx}^* \rangle$ and $\tau$ by relation (11) and employing them in relation (5), after using formula (12), enabled the nucleation rate $J(\langle t_{cx}^* \rangle)$ attained at a time corresponding to the average nucleation duration to be resolved explicitly. The shape of the thermal evolutions of $J(\langle t_{cx}^* \rangle)$ is well-known to experimenters. The evolutions attained here with the three interaction potentials are presented in Fig. 9. Each of them has the characteristic shape of the experimental curves (Ref. 25). It can be observed that the attraction has very little influence on the highest limit of the curves and that it broadens their plateau towards the high temperatures. The maximal value $7.10^{-7} C.\sigma_{Ar}^{-3}.t_{Ar}^{-1} = 7.10^{27} C.cm^{-3}.s^{-1}$ of the nucleation rate $J(\langle t_{cx}^* \rangle)$ at the average nucleation time, is shown in Fig. 9, is, considering the monoatomic nature of the liquid and the high pressure ($420.6 \; 10^5 Pa$) employed in the present numerical nucleations, in accordance with the order of magnitude of $10^{25} C.cm^{-3}.s^{-1}$ mentioned by Zarzycky in Ref. 25 about nucleation in atmospheric liquids of normal densities, in which the nucleation speed is high. It implies, in the present case, the quasi-inaccessibility, by fast quench, of deep (under $T^*=0.25$) super-cooling of liquids such as argon.

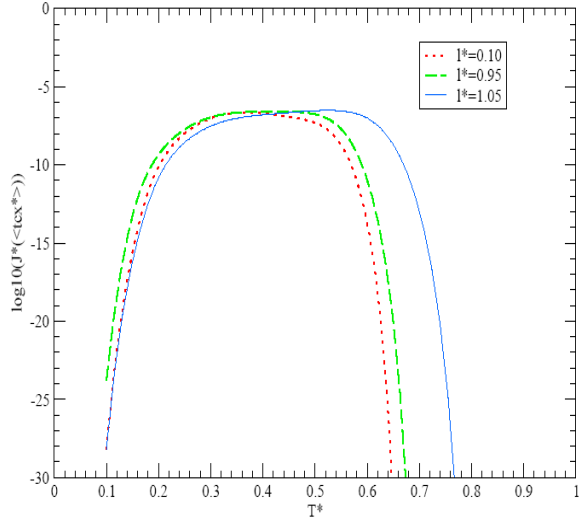

Figure 9. Forecasted evolutions of the realized nucleation rate

**M. Experimental validation**

The validity of the interfacial tension law (15)-correlations (17) set may be verified on argon by means of Fig. 7 and 9 relating to $l^*=0.95$. In the framework of low super-cooling and experimental durations in the order of a few minutes, Fig. 7 shows that super-cooling at 0.64 $T_{Ar}$ (77K) should yield an average nucleation duration of $10^{14} t_{Ar}$ (4mn) and Fig. 9 indicates that the nucleation rate should then be $1.8.10^{-17} C.\sigma_{Ar}^{-3}.t_{Ar}^{-1}$ ($2.12\,10^{23} C.cm^{-3}.s^{-1}$). Thus, the nucleation rate being stationary according to Fig. 8, it is by using formula (12) that these tests should enable verification of the validity of law (15).

**V. CONCLUSIONS**

The examination of CNT made in part II on the basis of the thermal evolutions of interfacial tension, carried out in molecular interaction conditions of fixed repulsion and varied attractions, has underlined an unclassifiable character of the evolutions inconsistent with the constant and regular behaviour of the other key sizes of nucleation. On the basis of this fact, the ambiguous status of interfacial tension has been resolved unfavorably for CNT, since interfacial tension could only keep its status of physical size to the detriment of the forecast accuracy of this theory.

The work described in part III to replace CNT targeted an alternative theory that keeps all its qualities of formal simplicity and efficiency. It again uses the CNT dimension analysis by adopting the sizes characteristic of nucleation as variables and retains the Zeldovich formula to globally describe incubation and free nucleation, for its simplicity and reliability to physical phenomenon. The expressions of time constant and stationary nucleation rate of this formula, without dimension anomaly, have been established and, in order to determine the universal constants they introduce, a mathematical form for the thermal law of interfacial tension has had to be proposed and minimal physical hypotheses, as realistic as possible, to be advanced. The remarkably accurate adjusting of this law to numerical simulations and the thermal behaviour of interfacial tension now constant and regular are no doubt signs of its physical relevance.

On the basis of this result, it is foreseeable that the obtained law will apply without change to all numerical or physical fluids of monoatomic particles. Concerning materials formed of small molecules, non reported testing computations performed on data from Ref. 26 related to nifedipine and felodipine showed that a new adjusting parameters determination should be necessary.

To finish, we will make the remark that, in this study, the resort to a large notion of measure has been performed three times: to detect the dimension anomalies, to criticize the thermal evolutions of interfacial tension in agreement with CNT and to establish a scales dependant interfacial tension thermal law. The specificity of Physics as an exact science indeed lies in that bridge between theory and experimentation, whether physical or numerical, which is the notion of measure with all its attributes: physical size and dimension.

**ACKNOWLEDGMENTS**

This work is the fruit of a freedom of research which can always be found inside the UMET in Lille 1 University. The authors would like to thank the French State for continuously and generously supporting its researchers.